\documentclass[11pt]{article}
\usepackage{amsmath,amsfonts,latexsym,graphicx,amssymb}
\date{}
\def\la{\langle\,}
\def\r{\,\rangle}
\newcommand{\eeq}{\end{eqnarray}}
\newcommand{\beq}{\begin{eqnarray}}

\newcommand{\mf}{\mathfrak{f}}

\newcommand{\bbR}{\mathbb{R}}
\newcommand{\bbC}{\mathbb{C}}

\def\con{{}_{\_\rule{-1pt}{0pt}\_}
\rule{-2pt}{0pt}\raise1.5pt\hbox{$\mid$}\hspace{2pt}}

\newtheorem{theorem}{Theorem}
\newtheorem{proposition}{Proposition}

\title{\bf Quantum Mechanics of Damped Systems II. Damping and
Parabolic Potential Barrier.}
\author{Dariusz Chru\'sci\'nski  \\
 Institute of Physics, Nicolaus Copernicus University \\
 ul. Grudzi\c{a}dzka 5/7, 87-100 Toru\'n, Poland}
\begin{document}

\maketitle

\begin{abstract}

We investigate the resonant states for the parabolic potential
barrier known also as inverted or reversed oscillator. They
correspond to the poles of  meromorphic continuation of the
resolvent operator to  the complex energy plane. As a byproduct we
establish an interesting relation between parabolic cylinder
functions (representing energy eigenfunctions of our system) and a
class of Gel'fand distributions used in our recent paper.
\end{abstract}

\vspace{.7cm}


\vspace{.3cm}


\numberwithin{equation}{section}


\section{Introduction}
\setcounter{equation}{0}

In a recent paper \cite{I} we have investigated a quantization of
the simple damped system\footnote{We slightly change the notation:
the coordinates $(x,p)$ used in \cite{I} are replaced by $(u,v)$
in the present paper.}
\begin{equation}  \label{dot-u}
\dot{u} = - \gamma u\ .
\end{equation}
To quantize this system we double the number of degrees of
freedom, i.e. together with (\ref{dot-u}) we consider $\dot{v} = +
\gamma v$. The enlarged system is a Hamiltonian one and its
quantization leads to the following quantum Hamiltonian:
\begin{equation}\label{H-uv}
  \widehat{H} = - \frac{\gamma}{2} (\widehat{u}\widehat{v} + \widehat{v}\widehat{u} )
  \ .
\end{equation}
We showed that the above system displays two families of
generalized eigenvectors $f^\pm_n$ corresponding to purely
imaginary eigenvalues $\widehat{H} f^\pm_n = \pm E_n f^\pm_n$.
These eigenvectors are interpreted as resonant states --- they
correspond to the poles of energy eigenfunctions when continued to
the complex energy plane. It turns out that resonant states are
responsible for the irreversible behavior. We showed that there
are two dense subspaces $\Phi_\pm \in L^2( \mathbb{R})$ such that
restriction of the unitary group $U(t) = e^{-i\widehat{H}t}$ to
$\Phi_\pm$ does no longer define a group but gives rise to two
semigroups: $U_-(t)=U(t)|_{\Phi_-}$ defined for $t\geq 0$ and
$U_+(t)=U(t)|_{\Phi_+}$ defined for $t\leq 0$. In the framework of
Gel'fand triplets (see e.g. \cite{Bohm-Gadella}) it means that the
quantum version of the damped system (\ref{dot-u}) corresponds to
the Gel'fand triplet:
\begin{equation}\label{}
  \Phi_- \, \subset \, L^2( \mathbb{R}) \, \subset \, \Phi_-'\ ,
\end{equation}
together with the Hamiltonian $\widehat{H}|_{\Phi_-}$. This system
serves as a simple example of Arno Bohm theory of resonances
\cite{Bohm} (see also \cite{RES-1,RES-2}) and illustrates
mathematical results of \cite{Gorini}.

In the present paper we continue to study this system but in a
different representation. Let us observe that  performing the
linear canonical transformation $(u,v) \longrightarrow (x,p)$:
\begin{equation}\label{CAN}
  u = \frac{ \gamma x - p}{\sqrt{2\gamma}}\ , \ \ \ \ \
  v = \frac{ \gamma x + p}{\sqrt{2\gamma}}\ ,
\end{equation}
one obtains for the Hamiltonian
\begin{equation}\label{H-qp}
\widehat{H} = \frac 12 ( \widehat{p}^2 - \gamma^2 \widehat{x}^2) \
.
\end{equation}
It represents the parabolic potential  barrier $V(x) = - \gamma^2
x^2/2$ and it was studied by several authors in various contexts
\cite{Kemble,Wheeler,Friedman,Ann1,Ann2,Castagnino,Shimbori1}. It
is well known that this system gives rise to the generalized
complex eigenvalues --- the physical reason for that is the
potential unbounded from below. We find the corresponding energy
eigenstates for (\ref{H-qp}). They are given in terms of parabolic
cylinder functions $D_\nu(x)$. Using the Gel'fand-Maurin spectral
decomposition we find the resolvent operator $R(z,\widehat{H}) =
(\widehat{H}-z)^{-1}$ and relate its poles to the resonant states.
As a byproduct we established a deep relation between the Gel'fand
distributions $u^\lambda_\pm$ \cite{Gel-Shi,Kanwal} (used in
\cite{I}) and parabolic cylinder functions $D_\nu(x)$. The details
are included in the Appendix.

\section{Inverted oscillator and complex eigenvalues}

Let us note that  $\widehat{H}$ defined in  (\ref{H-qp})
corresponds to the Hamiltonian of the harmonic oscillator with
purely imaginary frequency $\omega = \pm i\gamma$ (in the
literature it is also called an inverted or reversed oscillator).
The connection with a harmonic oscillator may be established by
the following scaling operator \cite{Kossak}:
\begin{equation}\label{V-lambda}
  \widehat{V}_\lambda := \exp\left( \frac{\lambda}{2} ( \widehat{x}\widehat{p}
  + \widehat{p}\widehat{x}) \right)\ ,
 \end{equation}
with $\lambda \in \mathbb{R}$. Using commutation relation
$[\widehat{x},\widehat{p}]=i$, this operator may be rewritten as
follows
\begin{equation}\label{}
 \widehat{V}_\lambda =  e^{-i\frac{\lambda}{2}}\, e^{\lambda\widehat{x}\widehat{p}}
=  e^{-i\frac{\lambda}{2}}\, e^{-i\lambda x\partial_x} \ ,
\end{equation}
and therefore it defines a complex dilation, i.e. the action of $
\widehat{V}_\lambda$ on a function $\varphi=\varphi(x)$ is given
by
\begin{equation}\label{}
\widehat{V}_\lambda\, \varphi(x) = e^{-i\frac{\lambda}{2}}\,
\varphi( e^{-i{\lambda}}\, x)\ .
\end{equation}
In particular one easily finds:
\begin{equation}\label{}
\widehat{V}_\lambda\, \widehat{x}\, \widehat{V}_\lambda^{-1} =
e^{-i\lambda} \widehat{x} \ , \hspace{1cm} \widehat{V}_\lambda\,
\widehat{p}\, \widehat{V}_\lambda^{-1} = e^{i\lambda} \widehat{p}
\ ,
\end{equation}
and hence
\begin{equation}\label{}
  \widehat{V}_\lambda\, ( \widehat{p}^2 - \gamma^2 \widehat{x}^2) \, \widehat{V}_\lambda^{-1}
   = e^{2i\lambda} \, ( \widehat{p}^2 - e^{-4i\lambda}\gamma^2 \widehat{x}^2)\ .
\end{equation}
Therefore, for $e^{4i\lambda} = -1$, i.e. $\lambda = \pm \pi/4$,
one has
\begin{equation}\label{}
\widehat{V}_{\pm \pi/4}\, \widehat{H} \, \widehat{V}_{\pm
\pi/4}^{-1} = \pm i \widehat{H}_{\rm ho}\ ,
\end{equation}
where
\begin{equation}\label{}
\widehat{H}_{\rm ho} =  \frac 12 ( \widehat{p}^2 + \gamma^2
\widehat{x}^2) \ ,
\end{equation}
stands for the oscillator Hamiltonian. In particular if $E^{\rm
ho}_n = \gamma (n + \frac 12)$ is an oscillator spectrum
\begin{equation}\label{}
\widehat{H}_{\rm ho} \psi^{\rm ho}_n = E^{\rm ho}_n \psi^{\rm
ho}_n \ ,
\end{equation}
then
\begin{equation}\label{}
\widehat{H} \mf^{\pm}_n = \pm E_n \mf^{\pm}_n \ ,
\end{equation}
with
\begin{equation}\label{En}
  E_n = iE^{\rm ho}_n = i\gamma \left( n + \frac 12 \right) \ ,
\end{equation}
and
\begin{equation}\label{psi-psi-ho}
  \mf^{\pm}_n(x) = \widehat{V}_{\mp \pi/4}\, \psi^{\rm ho}_n(x) = e^{\pm i \frac{\pi}{8}}
  \, \psi^{\rm ho}_n(e^{\pm i \frac{\pi}{4}}x) \ .
\end{equation}
Now, recalling that (see e.g. \cite{LL})
\begin{equation}\label{}
  \psi^{\rm ho}_n(x) = N_n \, e^{- \frac{\gamma}{2} x^2}
  H_n(\sqrt{\gamma}x)\ ,
\end{equation}
where $H_n$ stands for the n-th Hermite polynomial and the
normalization constant
\begin{equation}\label{Nn}
  N_n = \left( \frac{\sqrt{\gamma}}{2^n n!\sqrt{\pi}} \right)^{\frac 12}\ ,
\end{equation}
one obtains the following formulae for the generalized
eigenvectors of $\widehat{H}$:
\begin{equation}\label{psi-pm-n}
  \mf^{\pm}_n(x) = N^\pm_n \,  e^{ \mp i \frac{\gamma}{2} x^2}
  H_n(\sqrt{\pm i\gamma}x)\ ,
\end{equation}
with
\begin{equation}\label{Nn-pm}
  N^\pm_n = e^{\pm i \frac{\pi}{8}} \, N_n =
  \left( \frac{\sqrt{\pm i\gamma}}{2^n n!\sqrt{\pi}} \right)^{\frac 12}\
  .
\end{equation}
Clearly, $\mf^\pm_n$ are not elements from $L^2( \mathbb{R})$ but
they do belong to the dual of the Schwartz space ${\cal S}(
\mathbb{R}_x)'$, i.e. they are tempered distributions.

\begin{proposition} \label{PRO-psi}
Two families of generalized eigenvectors $\mf^\pm_n$ satisfy the
following properties:
\begin{enumerate}

\item they  are conjugated to each other:
\begin{equation}\label{f+--f-}
  \overline{\mf^+_n(x)} = \mf^-_n(x)\ ,
\end{equation}

\item they are orthonormal
\begin{equation}\label{}
  \la \mf^\pm_n | \mf^\mp_m\r = \delta_{nm}\ ,
\end{equation}

\item they are complete
\begin{equation}\label{}
  \sum_{n=0}^\infty \overline{\mf^\pm_n(x)}\, \mf^\mp_n(x') =
  \delta(x-x')\ .
\end{equation}
\end{enumerate}
\end{proposition}
The proof follows immediately from orthonormality and completness
of oscillator eigenfunctions $\psi^{\rm ho}_n$. Formula
(\ref{f+--f-}) implies that $\mf^+_n$ and $\mf^-_n$ are related by
the time reversal operator $\bf T$: ${\bf T}\psi :=
\overline{\psi}$. Recall \cite{I} that in $u$-representation  $\bf
T$ is unitary (it is defined by the Fourier transformation),
whereas in $x$-representation it is antiunitary.

\section{Change of representation}

It should be clear that there exists relation between generalized
eigenvectors $\mf^\pm_n(x)$ and $f^\pm_n(u)$ found in \cite{I}:
\begin{equation}\label{}
  f^+_n(u)\, \sim\, u^n\ , \hspace{1cm}   f^-_n(u) \, \sim\,
  \delta^{(n)}(u)\ .
\end{equation}
They define the same eigenvectors $\,|\pm n\r\,$ but in different
representations:
\[   \mf^\pm_n(x) = \la x|\pm n\r\ , \hspace{1cm}  f^\pm_n(u) = \la
u|\pm n\r \ . \]
To find this relation let us observe that the
canonical transformation (\ref{CAN}) is generated by the following
generating function
\begin{equation}\label{S-xu}
  S(x,u) = \frac{\gamma}{2} x^2 - \sqrt{2\gamma} xu +
  \frac{1}{2}u^2\ ,
\end{equation}
that is,
\begin{equation}\label{}
  p = \frac{\partial S}{\partial x} \ , \hspace{1cm} v =-
  \frac{\partial S}{\partial u} \ .
\end{equation}
Let us define a unitary operator
\[  {\cal U} \ : \ L^2( \mathbb{R}_u) \ \longrightarrow\ L^2(
\mathbb{R}_x)\ , \] by
\begin{equation}\label{U}
 f\ \ \ \longrightarrow\ \ \ ( {\cal U} f)(x) = C\,
  \int_{-\infty}^\infty f(u) e^{iS(x,u)}\, du\ ,
\end{equation}
 where the constant `$C$' is determined by
\begin{equation}\label{}
|C|^2\, \int_{-\infty}^\infty  e^{iS(x,u)}\, e^{-iS(x',u)}\, du =
\delta(x-x')\ .
\end{equation}
It implies $C=e^{i\alpha}\, C_0$, where $\alpha$ is an arbitrary
phase and
\begin{equation}\label{C0}
C_0 =   \left( \frac{\gamma}{2\pi^2} \right)^{\frac 14} \ ,
\end{equation}
In the next section it would be clear that a natural choice for
the phase is $\alpha=-\pi/8$. Clearly, $\cal U$ may be extended to
act on ${\cal S}( \mathbb{R}_u)'$. It is easy to show that
\begin{equation}\label{}
{\cal U} ( {\cal S}( \mathbb{R}_u)' ) \subset {\cal S}(
\mathbb{R}_x)'\  .
\end{equation}

\begin{proposition}   \label{PRO-psi-f}
The generalized eigenvectors $\mf^\pm_n \in {\cal S}(
\mathbb{R}_x)'$ and $f^\pm_n \in {\cal S}( \mathbb{R}_u)'$ are
related by:
\begin{equation}\label{}
  \mf^\pm_n = {\cal U}\,f^\pm_n\ .
\end{equation}
\end{proposition}
{\em Proof}. Let us show that $\mf^+_n = {\cal U} f^+_n$, that is
\begin{equation}\label{}
\mf^+_n(x)\, \sim\, \int u^n e^{iS(x,u)}du\ .
\end{equation}
Using the definition of $S(x,u)$ one has
\begin{eqnarray}\label{}
\int u^n e^{iS(x,u)}du &=& \left(-i\sqrt{2\pi \gamma}\right)^{-n}
e^{i\gamma \frac{x^2}{2}} \, \frac{d^n}{dx^n} \int e^{iu^2/2 -
i\sqrt{2\gamma}xu}du \nonumber
\\ &=&  \sqrt{-2\pi i} \left(-i\sqrt{2\pi \gamma}\right)^{-n} e^{-i\gamma \frac{x^2}{2}} \left( \,
e^{i\gamma {x^2}} \frac{d^n}{dx^n} e^{-i\gamma {x^2}} \right) \ .
\end{eqnarray}
Now, due to the well known formula for the Hermite polynomials
\begin{equation}\label{}
  e^{iz^2} \frac{d^n}{dz^n} e^{-iz^2} = (-1)^nH_n(z)\ ,
\end{equation}
one obtains
\begin{equation}\label{}
\int u^n e^{iS(x,u)}du = \sqrt{-2\pi i} \left(
\frac{i}{2}\right)^{-\frac n2} e^{-i\gamma \frac{x^2}{2}}
H_n(\sqrt{i\gamma}x)   \sim \mf^+_n(x)\ .
\end{equation}
To prove that $\mf^-_n = {\cal U} f^-_n$, let us note
that\footnote{It turns out that a function
\begin{equation*}\label{}
  \widetilde{S}(x,v) = - S(x,v) = - \frac{\gamma}{2} x^2 + \sqrt{2\gamma} xv
  -  \frac{1}{2}v^2\ ,
\end{equation*}
serves as a generating function for the canonical transformation
(\ref{CAN}):
\begin{equation*}\label{}
  p = \frac{\partial \widetilde{S}}{\partial x} \ , \hspace{1cm}  u=-
  \frac{\partial \widetilde{S}}{\partial v} \ .
\end{equation*}   }

\begin{equation}\label{}
  \mf^-_n(x) = \overline{\mf^+_n(x)} \, \sim\, \int u^n
  e^{-iS(x,u)}du\ .
\end{equation}
Now, taking into account that $f^+_n$ and $f^-_n$ are related by
the Fourier transformation
\begin{equation}\label{}
  u^n = \sqrt{2\pi} (-i)^n F^{-1}[ \delta^{(n)}(k)](u) \ ,
\end{equation}
one obtains
\begin{eqnarray}\label{}
\int u^n
  e^{-iS(x,u)}du &=&  \sqrt{2\pi} (-i)^n \int \delta^{(n)}(u)
  F^{-1}\left[ e^{-iS}\right](u)\, du\ .
\end{eqnarray}
Finally,
\begin{eqnarray}\label{}
F^{-1}\left[ e^{-iS}\right](u)  = \frac{1}{\sqrt{2\pi}} \int
e^{-iku}\, e^{-iS(x,k)}dk
= \sqrt{-i}\, e^{iS(x,u)}\ ,
\end{eqnarray}
and hence
\begin{equation}\label{}
\mf^-_n(x)\, \sim\, \int \delta^{(n)}(u) e^{iS(x,u)}du\ ,
\end{equation}
which ends the proof. \hfill $\Box$

\section{Energy eigenstates}

The spectrum of the self-adjoint operator (\ref{H-uv})  reads $
\sigma (\widehat{H} ) = (-\infty,\infty)$  and the corresponding
energy eigenstates (in $u$-representation) are given by (cf.
section~6 in \cite{I}):
\begin{equation}\label{}
  \psi^E_\pm(u) = \frac{1}{\sqrt{2\pi \gamma}} \, u_\pm^{-(iE/\gamma +
  1/2)}\ ,
\end{equation}
with $E \in \mathbb{R}$.  For the basic properties of the tempered
distributions $u_\pm^\lambda \in {\cal S}( \mathbb{R}_u)'$ we
refer the reader to \cite{Gel-Shi,Kanwal} (see also the Appendix
in \cite{I}). Now, using $(x,p)$ coordinates the corresponding
eigenvalue problem $\frac 12 (\widehat{p}^2 -
\gamma^2\widehat{x}^2)\chi^E=E\chi^E$ reads
\begin{equation}\label{}
  \partial^2_x \chi^E(x) + (\gamma^2x^2 + 2E)\chi^E(x) = 0\ .
\end{equation}
Introducing a new variable
\begin{equation}\label{}
  z = \sqrt{2i\gamma}\, x\ ,
\end{equation}
the above equation may be rewritten as follows
\begin{equation}\label{par-eq}
  \partial^2_z \chi^E + \left( \nu + \frac 12 - \frac{z^2}{4}
  \right) \chi^E = 0 \ ,
\end{equation}
with
\begin{equation}\label{nu-E}
  \nu = - \left( i \frac{E}{\gamma} + \frac 12 \right)\ ,
\end{equation}
which is the defining equation  for the parabolic cylinder
functions \cite{GR,Morse,AS}. Its solution $\chi^E(z)$ is  a
linear combination of $D_\nu(z)$, $D_\nu(-z)$, $D_{-\nu -1}(iz)$
and $D_{-\nu-1}(-iz)$.\footnote{These four functions are linearly
dependent. For the linear relation see e.g. formula 9.248 in
\cite{GR}.} On the other hand the energy eigenstates in
$x$-representation $\chi^E(x)$ may be obtained by applying the
operator $\cal U$ defined in (\ref{U}) to the corresponding
eigenstates in $u$-representation $\psi^E_\pm(u)$:
\begin{equation}\label{}
  \chi^E_\pm(x) =  ({\cal U}\psi^E_\pm)(x) = C
  \int_{-\infty}^\infty
  \psi^E_\pm(u)\, e^{iS(x,u)}\, du\ .
\end{equation}
Hence
\begin{eqnarray}\label{}
  \chi^E_+(x) &=& \frac{C}{\sqrt{2\pi \gamma}}\, e^{i\frac{\gamma}{2}x^2}
  \int_0^\infty u^\nu\, e^{-i\sqrt{2\gamma}xu + iu^2/2}\, du \nonumber
  \\ &=& \frac{C}{\sqrt{2\pi \gamma}}\, \sqrt{i}^{\nu + 1}\, e^{-\frac{y^2}{4}}
  \int_0^\infty \xi^\nu\, e^{y\xi - \xi^2/2}\, d\xi\ ,
\end{eqnarray}
with $y= \sqrt{-2i\gamma}\,x$, and using an integral
representation for $D_p(y)$ (formula 9.241(2) in
\cite{GR}):\footnote{The validity of this formula is restricted in
\cite{GR} for ${\rm Re}\, p < 0$. However, as we shall show (see
the proof of Proposition~\ref{PRO-Dp}), it is valid for all $p \in
\mathbb{C}$.}
\begin{equation}\label{}
  D_p(y) = \frac{e^{-\frac{y^2}{4}}}{\Gamma(-p)}\,
  \int_0^\infty \xi^{-p-1}\, e^{-y\xi - \xi^2/2}\, d\xi\ ,
\end{equation}
one finds
\begin{equation}\label{chi-E+}
  \chi^E_+(x) = \frac{C_0}{\sqrt{2\pi \gamma}}\, \sqrt{i}^{  \nu +
  \frac 12}\, \Gamma(\nu + 1)
  D_{-\nu - 1}(-\sqrt{-2i\gamma}x)\ ,
\end{equation}
with $\nu$ given in (\ref{nu-E}). Similarly, using an obvious
relation  $(-u)^\lambda_+ = u^\lambda_-$, one obtains:
\begin{equation}\label{chi-E-}
 \chi^E_-(x) = \frac{C_0}{\sqrt{2\pi \gamma}}\, \sqrt{i}^{\ \nu +
  \frac 12}\, \Gamma(\nu + 1)
  D_{-\nu - 1}(\sqrt{-2i\gamma}x)\ ,
\end{equation}
that is, $\chi^E_-(x) = \chi^E_+(-x)$. Actually, instead of
$\chi^E_\pm$ one may use energy  eigenstates with the definite
parity:
\begin{eqnarray}
\chi^E_{\rm even} &=& \frac{1}{\sqrt{2}} \left( \chi^E_+ + \chi^E_- \right) \ ,\\
\chi^E_{\rm odd} &=& \frac{1}{\sqrt{2}} \left( \chi^E_+ - \chi^E_-
\right) \ ,
\end{eqnarray}
that is,
\begin{equation}
{\bf P}\, \chi^E_{\rm even} = \chi^E_{\rm even}\ , \hspace{1cm}
 {\bf P}\, \chi^E_{\rm odd} = - \chi^E_{\rm odd}\ ,
\end{equation}
where $\bf P$ stands for the parity operator.

\begin{proposition}
Energy eigenstates $\chi^E_\pm$ satisfy:
\begin{equation}\label{}
  \int_{-\infty}^\infty \overline{\chi^E_\pm(x)}\chi^{E'}_\pm(x)\, dx
  = \delta(E-E')\ ,
\end{equation}
and
\begin{equation}\label{}
  \int_{-\infty}^\infty \overline{\chi^E_\pm(x)}\chi^E_\pm(x')\, dE
  = \delta(x-x')\ .
\end{equation}
\end{proposition}
The proof follows immediately from the analogous properties
satisfied by energy eigenstates $\psi^E_\pm$ in $u$-representation
\cite{I}.

In \cite{I} we have used also another generalized basis
$F[\psi^{-E}_\pm](u)$. Now, we find its $\cal U$ image  in ${\cal
S}( \mathbb{R}_x)'$. Recalling the Fourier transformation of
$x^\lambda_\pm$ (see \cite{Gel-Shi} and Appendix in \cite{I}):
\begin{equation}\label{}
  F[x^\lambda_\pm](u) = \frac{\pm i}{\sqrt{2\pi}}\, e^{\pm
  i\lambda\frac{\pi}{2}}\, \Gamma(\lambda+1)(u+i0)^{-\lambda-1}\ .
\end{equation}
one has
\begin{equation}\label{}
  F[\psi^{-E}_+](u) = \frac{1}{\sqrt{2\pi\gamma}} \,
  \frac{(-i)^\nu}{\sqrt{2\pi}}\, \Gamma(-\nu) (u+i0)^\nu\ .
\end{equation}
Therefore, the corresponding $x$-representation
\begin{equation}\label{}
\eta^E_+(x) = ({\cal U}\, F[\psi^{-E}_+])(x)\ ,
\end{equation}
is given by
\begin{eqnarray}\label{}
  \eta^E_+(x) &=& \frac{C}{\sqrt{2\pi\gamma}} \,
  \frac{(-i)^\nu}{\sqrt{2\pi}}\, \Gamma(-\nu)\,
  \int_{-\infty}^\infty (u+i0)^\nu e^{iS(x,u)} du \nonumber\\ &=&
\frac{C}{\sqrt{2\pi\gamma}} \,
  \frac{(-i)^\nu}{\sqrt{2\pi}}\, (2\sqrt{i})^{\nu+1}\, \Gamma(-\nu)\,
e^{\frac{y^2}{4}}  \int_{-\infty}^\infty (\xi+i0)^\nu\, e^{-2\xi^2
-2iy\xi}\, d\xi\ ,
\end{eqnarray}
with $y= \sqrt{2i\gamma}x$. Now, using the following integral
representation (formula 9.241(1) in \cite{GR})
\begin{equation}\label{}
  D_\nu(y) = \frac{1}{\sqrt{\pi}}\, 2^{\nu + \frac 12} (-i)^\nu
  e^{\frac{y^2}{4}}\, \int_{-\infty}^\infty (\xi+i0)^\nu\, e^{-2\xi^2
+2iy\xi}\, d\xi\ ,
\end{equation}
one obtains
\begin{equation}\label{eta-E+}
  \eta^E_+(x) =  \frac{C_0}{\sqrt{2\pi\gamma}} \,
  {\sqrt{i}}^{\ \nu+ \frac 12}\, \Gamma(-\nu) D_\nu(-\sqrt{2i\gamma}x)\ .
\end{equation}
Similarly one shows that
\begin{equation}\label{}
\eta^E_-(x) = ({\cal U}\, F[\psi^{-E}_-])(x)\ ,
\end{equation}
is given by
\begin{equation}\label{eta-E-}
  \eta^E_-(x) =  \frac{C_0}{\sqrt{2\pi\gamma}} \,
  {\sqrt{i}}^{\ \nu+\frac 12}\, \Gamma(-\nu) D_\nu(\sqrt{2i\gamma}x)\
  .
\end{equation}
Let us note, that
\begin{equation}\label{}
  \overline{\nu +1} = -\nu\ ,
\end{equation}
and
\begin{equation}\label{}
  \overline{{\sqrt{i}}^{\ \nu+\frac 12}} = {\sqrt{i}}^{\ \nu + \frac 12} \ .
\end{equation}
Clearly, the transition $\nu+1 \longrightarrow -\nu$ is equivalent
to $E \longrightarrow -E$ and  it corresponds to the fact that
$\widehat{H} \eta^E_+ = -E \eta^E_+$ while $\widehat{H} \chi^E_+ =
+E \chi^E_+$. The symmetry between $\chi^E_\pm$ and $\eta^E_\pm$
fully justifies the specific choice of the phase factor in the
constant $C$. One has
\begin{equation}\label{eta-chi}
\eta^E_\pm(x) = \overline{\chi^E_\pm(x)}\ ,
\end{equation}
that is they are related by the time reversal operator $\bf T$:
$\eta^E_\pm = {\bf T}\, \chi^E_\pm$. Thus energy eigenstates
$\eta^E_\pm$ correspond to the time reversed system.
 This way all four solutions of (\ref{par-eq}) were used to
construct four families of energy eigenstates: $\chi^E_+$,
$\chi^E_-$, $\eta^E_+$ and $\eta^E_-$.

\section{Analytic continuation, resolvent and resonances}

Now, let us continue the energy eigenfunctions $\chi^E_\pm$ and
$\eta^{E}_\pm$ into the energy complex plane $E \in \bbC$ and let
us study its analyticity as functions of $E$.

\begin{proposition}  \label{PRO-Dp}
The parabolic cylinder function $D_\lambda(z)$ is an analytic
function of $\lambda\in \mathbb{C}$.
\end{proposition}
For the proof see the Appendix. Due to the above proposition the
analytic properties of the energy eigenfunctions are entirely
governed by the analytic properties of the $\Gamma$ function which
is present in the definition of $\chi^E_\pm$ and $\eta^E_\pm$.
Since $\Gamma(\lambda)$ has simple poles at $\lambda=-n$, with
$n=0,1,2,\ldots$, functions $\chi^E_\pm$ have poles at $E=-E_n$,
whereas functions $\eta^E_\pm$ have poles at $E=E_n$, where $E_n$
is defined in (\ref{En}). Using a well known formula for a residue
of the $\Gamma$ function
\begin{equation}\label{}
  {\rm Res}\left(\Gamma(\lambda);\lambda=-n\right) = \frac{(-1)^n}{n!}\ ,
\end{equation}
one has
\begin{equation}\label{}
  {\rm Res}\left(\chi^E_\pm(x);-E_n\right) = \frac{C_0}{\sqrt{2\pi\gamma}}\,
  \frac{(-1)^n}{n!}\, \sqrt{i}^{\ -n - \frac 12}\, D_n(\mp\sqrt{-2i\gamma}x)\ ,
\end{equation}
and
\begin{equation}\label{}
  {\rm Res}\left(\eta^E_\pm(x);+E_n\right) = \frac{C_0}{\sqrt{2\pi\gamma}}\,
\frac{(-1)^n}{n!}\, \sqrt{i}^{\ n + \frac 12}\,
D_n(\mp\sqrt{2i\gamma}x)\  .
\end{equation}
Hence,  using the  relation \cite{GR,Morse,AS}:\footnote{In
\cite{GR} the corresponding equation 9.253 has a wrong sign.}
\begin{equation}\label{}
  D_n(z) = 2^{- \frac n2}\, e^{-\frac{z^2}{4}}\, H_n\left(
  \frac{z}{\sqrt{2}}\right)\ ,\ \hspace{1cm} n=0,1,2,\ldots\ ,
\end{equation}
together with
\begin{equation}\label{}
  H_n(-z) = (-1)^nH_n(z)\ ,
\end{equation}
one obtains
\begin{equation}\label{}
 {\rm Res}\left(\chi^E_\pm(x);-E_n\right) \, \sim\,  \mf^+_n(x)\ ,
\end{equation}
and
\begin{equation}\label{}
 {\rm Res}\left(\eta^E_\pm(x);+E_n\right) \, \sim\,  \mf^-_n(x)\ .
\end{equation}
Now, it is natural to introduce two Hardy classes of functions
\cite{Duren}. Recall, that a smooth function $f=f(E)$ is in the
Hardy class from above ${\cal H}^2_+$ (from below ${\cal H}^2_-$)
if $f(E)$ is a boundary value of an analytic function  in the
upper, i.e. $\mbox{Im}\, E\geq 0$ (lower, i.e. $\mbox{Im}\, E\leq
0$) half complex $E$-plane vanishing faster than any power of $E$
at the upper (lower) semi-circle $|E| \rightarrow \infty$. Define
\begin{equation}
\Phi_- := \Big\{ \phi \in {\cal S}( \mathbb{R}_x)\, \Big| \,
f(E):= \la \chi^E_\pm | \phi \r \in {\cal H}^2_-\, \Big\} \ ,
\end{equation}
and
\begin{equation}
\Phi_+ := \Big\{ \phi \in {\cal S}( \mathbb{R}_x)\, \Big| \,
f(E):=\la \eta^{E}_\pm | \phi \r \in {\cal H}^2_+\, \Big\} \ .
\end{equation}
It is evident from (\ref{eta-chi}) that $  \Phi_+ =
\overline{\Phi_-}$, that is
\begin{equation}\label{}
  \Phi_+ = {\bf T}({\Phi_-})\ .
\end{equation}
 Due  to the
Gel'fand-Maurin spectral theorem \cite{RHS1,RHS2} any function
$\phi^-\in \Phi_-$ may be decomposed with respect to $\chi^E_\pm$
family
\begin{equation} \label{GM-1}
\phi^-(x) =    \sum_\pm  \int_{-\infty}^\infty dE\, \chi^E_\pm(x)
\la \chi^E_\pm|\phi^-\r \ ,
\end{equation}
and any function $\phi^+\in \Phi_+$  may be decomposed with
respect to $\eta^E_\pm$ family
\begin{equation} \label{GM-2}
\phi^+(x) =    \sum_\pm \int_{-\infty}^\infty dE\, \eta^E_\pm(x)
\la \eta^E_\pm|\phi^+\r \ .
\end{equation}
Applying the Residue Theorem one easily proves the following

\begin{theorem} For any function $\phi^\pm \in \Phi_\pm$ one has
\beq  \label{phi-} \phi^-(x) = \sum_{n=0}^\infty \mf^-_n(x) \la
\mf^+_n|\phi^-\r \ , \eeq and \beq   \label{phi+} \phi^+(x) =
\sum_{n=0}^\infty \mf^+_n(x) \la \mf^-_n|\phi^+\r \ . \eeq
\end{theorem}
The proof goes along the same lines as the corresponding proof of
Theorem~2 in \cite{I}. The above theorem implies the following
spectral resolutions of the Hamiltonian:
\begin{equation} \label{GMH-1}
\widehat{H} =  \sum_\pm \int_{-\infty}^\infty dE\, E | \chi^E_\pm
\r \la \chi^E_\pm| = - \sum_{n=0}^\infty E_n
|\mf^-_n\r\la\mf^+_n|\ ,
\end{equation}
on $\Phi_-$, and
\begin{equation} \label{GMH-2}
\widehat{H} =  \sum_\pm \int_{-\infty}^\infty dE\, E | \eta^E_\pm
\r \la \eta^E_\pm| =   \sum_{n=0}^\infty E_n
|\mf^+_n\r\la\mf^-_n|\ ,
\end{equation}
on $\Phi_+$. The same techniques may be applied for the resolvent
operator
\begin{equation}\label{}
  R(z,\widehat{H}) = \frac{1}{\widehat{H}-z}\ .
\end{equation}
One obtains
\begin{equation}\label{}
R(z,\widehat{H})  = \sum_\pm \int_{-\infty}^\infty\,
\frac{dE}{E-z}\, | \chi^E_\pm \r \la \chi^E_\pm| \,= \,
\sum_{n=0}^\infty\, \frac{1}{-E_n-z} \, |\mf^-_n\r\la\mf^+_n|\ ,
\end{equation}
on $\Phi_-$, and
\begin{equation}\label{}
R(z,\widehat{H})  = \sum_\pm \int_{-\infty}^\infty\,
\frac{dE}{E-z} \, | \eta^E_\pm \r \la \eta^E_\pm|\, =\,
\sum_{n=0}^\infty \, \frac{1}{E_n-z}\, |\mf^+_n\r\la\mf^-_n|\ ,
\end{equation}
on $\Phi_+$. Hence, $R(z,\widehat{H})|_{\Phi_-}$ has poles at
$z=-E_n$, and $R(z,\widehat{H})|_{\Phi_+}$ has poles at $z=E_n$.
As usual eigenvectors $\mf^-_n$ and $\mf^+_n$ corresponding to
poles of the resolvent  are interpreted as resonant states. Note,
that
\begin{equation}\label{}
  - \frac{1}{2\pi i} \oint_{\gamma_n} R(z,\widehat{H})dz =
  |\mf^+_n\r\la\mf^-_n| := \widehat{P}_n \ ,
\end{equation}
where $\gamma_n$ is a closed curve that encircles the singularity
$z=E_n$. Clearly,
\begin{equation}\label{}
  \widehat{P}_n \cdot \widehat{P}_m = \delta_{nm}\widehat{P}_n\ ,
\end{equation}
and the spectral decomposition of $\widehat{H}$ may be written as
follows:
\begin{equation}\label{}
  \widehat{H} =  \sum_{n=0}^\infty \, E_n \widehat{P}_n = - \sum_{n=0}^\infty
  \, E_n \widehat{P}_n^\dag\ .
\end{equation}

Finally, let us note, that restriction of the unitary group
$U(t)=e^{-i\widehat{H}t}$ defined on the Hilbert space $L^2(
\mathbb{R})$ to $\Phi_\pm$ no longer defines a group. It gives
rise to two semigroups:
\begin{equation}
 U_-(t) \ :\ \Phi_-
\ \longrightarrow\ \Phi_-\ ,\ \ \ \ \ \ {\rm for}\ \ \ t\geq 0\ ,
\end{equation}
and
\begin{equation}
  U_+(t) \ :\ \Phi_+ \ \longrightarrow\ \Phi_+\ ,\ \ \ \ \ \
{\rm for}\ \ \ t\leq 0\ .
\end{equation}
Using (\ref{GMH-1}), (\ref{GMH-2}) and the formula for
$E_n=i\gamma (n +\frac 12)$ one finds:
\begin{eqnarray} \label{phi:-}
\phi^-(t) = U_-(t)\phi^- = \sum_{n=0}^\infty e^{-\gamma(n+ \frac
12)t} \, \widehat{P}^\dag_n\phi^- \ ,
\end{eqnarray}
for $t\geq 0$, and
\begin{eqnarray} \label{phi:+}
\phi^+(t) = U_+(t)\phi^+ = \sum_{n=0}^\infty e^{\gamma(n+ \frac
12)t} \, \widehat{P}_n\phi^+ \ ,
\end{eqnarray}
for $t\leq 0$.  We stress that $\phi^-_t$ ($\phi^+_t$) does belong
to $L^2(\bbR)$ also for $t<0$ ($t>0$). However, $\phi^-_t \in
\Phi_-$ ($\phi^+_t \in \Phi_+$) only for $t\geq 0$ ($t\leq 0$).
This way the irreversibility enters the dynamics of the reversed
oscillator by restricting it to the dense subspace $\Phi_\pm$ of
$L^2(\bbR)$.


\section{Scattering vs. resonant states}

To compare the physical properties of energy eigenstates
$\chi^E_\pm$ and $\eta^E_\pm$ and resonant states $\mf^\pm_n$ let
us investigate its asymptotic behavior at $x \longrightarrow \pm
\infty$. Following \cite{AS}
 (see also \cite{Ann1,Ann2}) one finds\footnote{Putting $a= - E/\gamma$ in equation 19.17.9 in
\cite{AS} and using  relation 19.3.1
\begin{equation*}\label{}
  U(a,x) = D_{-a-\frac 12}(x)\ ,
\end{equation*}
one finds:
\begin{equation*}\label{}
  U\left( - i \frac{E}{\gamma}, \sqrt{2\gamma}xe^{-\frac 14 i\pi}\right) =
  D_{-\nu - 1}(\sqrt{-2i\gamma}x) \, \sim \, \chi^E_-(x)\ .
\end{equation*}  }
\begin{equation}\label{}
\chi^E_-(x \rightarrow +\infty) \, \sim \, \sqrt{\frac{1}{x}}\,
\exp\left[ i \left( \frac{\gamma}{2}x^2 + \frac{E}{\gamma}
\log(\sqrt{2\gamma}x) + \frac{\pi}{4}\frac{E}{\gamma} +
\frac{\pi}{8} \right)\right] \ ,
\end{equation}
and
\begin{eqnarray}\label{}
\chi^E_-(x \rightarrow -\infty)  &\sim& i \sqrt{\frac{1}{x}}
\left\{ \left(1 + e^{-2\pi\frac{E}{\gamma}}\right) \exp\left[ -i
\left( \frac{\gamma}{2}x^2 + \frac{E}{\gamma}
\log(\sqrt{2\gamma}x) - \frac{\pi}{4}\frac{E}{\gamma} +
\frac{3\pi}{8} +\phi \right)\right] \right.\nonumber \\  &-&
\left. e^{-\pi \frac{E}{\gamma}} \exp\left[ i \left(
\frac{\gamma}{2}x^2 - \frac{E}{\gamma} \log(\sqrt{2\gamma}x) -
\frac{\pi}{4}\frac{E}{\gamma} + \frac{\pi}{8}  \right)\right]
\right\}\ ,
\end{eqnarray}
where $\phi = {\rm arg}\, \Gamma(-i \frac{E}{\gamma} + \frac 12) =
\Gamma(\nu +1)$. Hence energy eigenstates $\chi^E_-$ represent
scattering states (see \cite{Ann1} for more details). The same is
true for $\chi^E_+$ and $\eta^E_\pm$. In particular one finds for
the reflection and transmission amplitudes $R$ and $T$ for
$\chi^E_\pm$ scattering states \cite{Wheeler,Ann1}:
\begin{eqnarray}\label{}
R(\chi^E_\pm) &=& - \frac{i}{\sqrt{2\pi}}\, e^{-\frac{\pi
E}{2\gamma}} \, \Gamma\left(\frac 12 - i\frac{E}{\gamma}\right)\ ,
\\ T(\chi^E_\pm) &=&  \frac{1}{\sqrt{2\pi}}\, e^{\frac{\pi
E}{2\gamma}} \, \Gamma\left(\frac 12 - i\frac{E}{\gamma}\right)\ .
\end{eqnarray}
Clearly, computing $R$ and $T$ for time-reversed $\eta^E_\pm$
scattering states one finds:
\begin{equation}\label{}
R(\eta^E_\pm) = \overline{ R(\chi^E_\pm)} \ , \hspace{1cm}
T(\eta^E_\pm) = \overline{ T(\chi^E_\pm)}\ .
\end{equation}
Note, that $R(\chi^E_\pm)$ and $ T(\chi^E_\pm)$ have poles at
$E=-E_n$, whereas $R(\eta^E_\pm)$ and $ T(\eta^E_\pm)$ have poles
at $E=+E_n$. Obviously, the corresponding reflection and
transition coefficients $|R|^2$ and $|T|^2$ are time-reversal
invariant.

On the other hand the eigenstates $\mf^\pm_n$ behave as follows:
\begin{equation}\label{}
  \mf^+_n(x \rightarrow \pm\infty) \, \sim \, (\pm\sqrt{i\gamma}x)^n\,
  e^{-i\frac{\gamma}{2}x^2}\ ,
\end{equation}
and
\begin{equation}\label{}
  \mf^-_n(x \rightarrow \pm\infty) \, \sim \, (\pm\sqrt{-i\gamma}x)^n\,
  e^{i\frac{\gamma}{2}x^2}\ .
\end{equation}
Note, that $\mf^-_n$ are purely outgoing states, whereas $\mf^+_n$
are purely ingoing states. Moreover, resonant states have
Breit-Wigner energy distribution. Indeed,
\begin{equation}\label{}
  \la \chi^E_-| \mf^+_n \r\, \sim\,  \Gamma(-\nu)\, \int_{-\infty}^\infty
 D_\nu(\sqrt{2\gamma i}x)\,\mf^+_n(x)\,dx\ .
\end{equation}
Now, $D_\nu$ is an entire function of $\nu$ and $\Gamma(-\nu)$ has
poles at $\nu = k \in \mathbb{N}$. In the domain where $n+1 >
\mbox{Re}\, \nu \geq 1$ one has
\begin{equation}\label{}
  \Gamma(-\nu) = {\rm analytical\ part}\, +\, \sum_{k=0}^n
  \frac{(-1)^k}{k!(k-\nu)^k} \ .
\end{equation}
Hence,
\begin{eqnarray}\label{}
\la \chi^E_-| \mf^+_n \r & \sim & {\rm analytical\ function\ of}\
E \, + \, \sum_{k=0}^n \frac{(-1)^k}{k!\left(k + i
\frac{E}{\gamma} +
\frac 12\right)} \, \la \mf^-_k|\mf^+_n \r \nonumber \\
& \sim& {\rm analytical\ function\ of}\ E \, + \frac{\gamma}{E -
E_n}\ ,
\end{eqnarray}
which is consistent with the Breit-Wigner formula.

\appendix

\section{Appendix}
\label{PROOFS}

The integral formula 9.241(2) in \cite{GR}
\begin{equation}\label{D-p}
  D_\lambda(y) = \frac{e^{-\frac{y^2}{4}}}{\Gamma(-\lambda)}\,
  \int_{-\infty}^\infty \xi^{-\lambda-1}_+\, e^{-y\xi - \xi^2/2}\, d\xi\ ,
\end{equation}
contains two objects: $\Gamma(-\lambda)$ and a distribution
$\xi^{-\lambda-1}$ which are singular for $\lambda=0,1,2,\ldots .$
However, it is easy to see \cite{Gel-Shi} that
\begin{equation}\label{}
  \frac{\xi^{-\lambda-1}_+}{\Gamma(-\lambda)}\Bigg|_{p=n} = \delta^{(n)}(\xi)\
  ,
\end{equation}
which shows that (\ref{D-p}) defines an entire function of
$\lambda\in \mathbb{C}$. The same is true for
\begin{equation}\label{D-p-}
  D_\lambda(y) = \frac{e^{-\frac{y^2}{4}}}{\Gamma(-\lambda)}\,
  \int_{-\infty}^\infty \xi^{-\lambda-1}_-\, e^{y\xi - \xi^2/2}\, d\xi\ ,
\end{equation}
due to
\begin{equation}\label{}
  \frac{\xi^{-\lambda-1}_-}{\Gamma(-\lambda)}\Bigg|_{\lambda=n} = (-1)^n\delta^{(n)}(\xi)\
  .
\end{equation}
The second integral representation given by 9.241(1) in \cite{GR}
\begin{equation}\label{D-p2}
  D_\lambda(y) = \frac{1}{\sqrt{\pi}}\, 2^{\lambda + \frac 12} (-i)^\lambda
  e^{\frac{y^2}{4}}\, \int_{-\infty}^\infty (\xi+i0)^\lambda\, e^{-2\xi^2
+2iy\xi}\, d\xi\ ,
\end{equation}
where $(\xi + i0)^\lambda = \xi^\lambda_+ +
e^{i\pi\lambda}\xi^\lambda_-$, seems to have poles at $\lambda
=-1,-2,\ldots .$ However, the limit $
  \lim_{\lambda \rightarrow -n} (\xi + i0)^\lambda$ is well
  defined \cite{Gel-Shi}
\begin{equation}\label{}
(\xi + i0)^{-n} = \xi^{-n} - \frac{i\pi (-1)^{n-1}}{(n-1)!}\,
\delta^{(n-1)}(\xi)\ .
\end{equation}
Thus, formula (\ref{D-p2}) also defines an entire function of
$\lambda$.

\section*{Acknowledgments}

I would like to  thank
 Andrzej Kossakowski  for many
interesting and stimulating discussions.

\end{document}